\documentclass[prl,twocolumn,superscriptaddress]{revtex4}

\usepackage[dvips]{graphics}
\usepackage{amsmath, amsthm, amssymb}
\usepackage{graphicx}
\usepackage{dcolumn}
\usepackage{bm}

\newcommand{\de}[1]{\left( #1 \right)}
\newcommand{\De}[1]{\left[ #1 \right]}
\newcommand{\DE}[1]{\left\{ #1 \right\}}

\newcommand{\ket}[1]{\left| #1 \right\rangle}
\newcommand{\bra}[1]{\left\langle #1 \right|}

\newcommand{\abs}[1]{\left| #1 \right|}

\newcommand{\tr}{\mathrm{Tr}}

\newcommand{\etal}{{\it{et al.}}}

\newtheorem{prop}{Proposition}

\begin{document}

\title{Geometrically induced singular behavior of entanglement}
\author{D. Cavalcanti}
\affiliation{ICFO-Institut de Ciencies Fotoniques, Mediterranean Technology Park, 08860 Castelldefels (Barcelona), Spain}

\author{P. L. Saldanha}
\author{O. Cosme}
\affiliation{Departamento de F\'{\i}sica, Universidade Federal de
Minas Gerais, Caixa Postal 702, 30123-970, Belo Horizonte, MG,
Brazil}

\author{F. G. S. L. Brand\~ao}
\affiliation{QOLS, Blackett Laboratory, Imperial College
London, London SW7 2BW, UK}
\affiliation{Institute for Mathematical Sciences, Imperial
College London, London SW7 2BW, UK}
\author{C. H. Monken}
\author{S. P\'adua}
\author{M. Fran\c ca Santos}
\affiliation{Departamento de F\'{\i}sica, Universidade
Federal de Minas Gerais, Caixa Postal 702, 30123-970, Belo
Horizonte, MG, Brazil}
\author{M. O. Terra Cunha}
\affiliation{Departamento de Matem\'atica, Universidade
Federal de Minas Gerais, Caixa Postal 702, 30123-970, Belo
Horizonte, MG, Brazil}

\begin{abstract}
We show that the geometry of the set of quantum states
 plays a crucial role in the behavior
of entanglement in different physical systems. More specifically it
is shown that singular points at the border of the set of
unentangled states appear as singularities in the dynamics of
entanglement of smoothly varying quantum states. We illustrate this result by implementing
a photonic parametric down conversion experiment.
Moreover, this effect is connected to
recently discovered singularities in
condensed matter models.
\end{abstract}

 \maketitle

Entanglement, a genuine quantum correlation, plays a crucial role
in different physical situations ranging from information
processing \cite{Eke} to quantum many-particle phenomena
\cite{RevManyBdy}. Similarly to thermodynamics, smooth variations
of controllable parameters which characterize a physical system
may lead to singular behavior of entanglement quantifiers. In some
cases, in similarity to quantum phases transitions
\cite{ON02-QPT}, these singularities are attested by abrupt
changes in the quantum state describing the system. However,
unexpected singularities may appear even when the quantum state
varies smoothly \cite{Osterloh-Roscilde1-Roscilde2}. Here we
demonstrate how the geometry of the set of unentangled states can
be related to singular behavior in physical phenomena. In
particular we show that singularities at the boundary of this set
can be detected by measuring the amount of entanglement of
smoothly varying quantum states.

Entangled states are defined as the states of composed quantum
systems which cannot be written as a convex sum of products of
density matrices for each composing part \cite{Werner}. Separable
states, on the other hand, admit such a representation and form a
convex, closed set with positive volume (for finite dimensional
systems) \cite{Sanpera}. This set, henceforth designated by $S$,
is a subset of $D$, the set of all density matrices ($S\subset
D$), which is also convex and closed. Therefore, a natural
geometric way to quantify entanglement is to see how far - using
some definition of distance on the state space - an entangled
state is from the set $S$. This has been carried over for a
variety of notions of distance, generating different measures of
entanglement \cite{HHHH-MartinShash}. One of these geometric
quantifiers is the random robustness, $R_R$, defined for any state
$\rho$ as the minimum $s$ ($s\ge0$) such that the state
\begin{equation}
\sigma=\frac{\rho+s \pi}{1+s} \label{rob}
\end{equation}
is separable ($\pi = I/d$, where $I$ is the identity matrix and $d$ the total dimension
of the state space) \cite{rob}. The physical motivation is clear:
$\sigma$ represents a mixture of $\rho$ with the random state $\pi$,
and $R_R\de{\rho}$ quantifies how much of this noise must be added to
$\rho$ in order to obtain a separable state.

\begin{figure}[ht]\centering
           \includegraphics[scale=0.26]{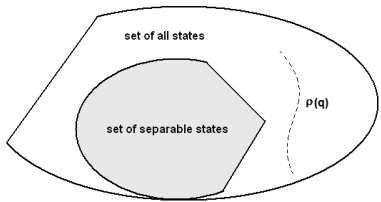}\\
          \caption{{\bf State space.} The dotted line represents the path $\rho(q)$
  followed by $\rho$ when parameter $q$ is changed. It is worth
  noting that $S$ can present singular points in its shape and to remember that the
  ``true'' picture is much subtler, given the large dimensionality of even
  the simplest example \cite{Zyc1}.}
  \label{figGPT}
\end{figure}

The main result of this Letter is to show that $R_R$ can be used
to investigate the shape of the boundary of $S$, $\partial S$. The
principle is to take an entangled state depending smoothly on one
parameter $q$ and compute $R_R$ as a function of $q$. The
one-parameter-dependent density matrices $\rho(q)$ can be seen as
a curve in the set of quantum states as shown in
Fig.~\ref{figGPT}. Singularities at $\partial S$ will show up as
singularities in
$R_R\de{\rho(q)}$. 
This statement is general for any finite dimension and will be
formalized by the contrapositive: if $\partial S$ is non-singular, then
$R_R\de{\rho(q)}$ is also non-singular. More precisely:
\begin{prop}
\label{teo} Let $D$ be a closed, convex set. Let $S\subset D$ also
be closed and convex, with $\pi$ a point in the interior of $S$. If
$\partial S$ is a $C^m$ manifold  and the states $\rho(q)$ describe
a $C^m$ curve in $D$ with no points in the interior of $S$ and
obeying the condition that the tangent vector $\rho'\de{q}$ is never
parallel to $\pi - \rho\de{q}$, then $R_R (\rho(q))$ is also a $C^m$
function.
\end{prop}
One must remember that a manifold is called $C^m$ if it can be
parameterized by functions with continuous derivatives up to order
$m$ ~\cite{Spi}. The reader can change $C^m$ by smooth, in the usual
sense of $C^{\infty}$, with almost no loss (actually, we use smooth throughout this Letter
in the less precise sense of ``as regular as necessary''). Other topological remarks
before the proof:
the fact that $S$ has interior points implies that $S$ and $D$ have
the same dimensionality (since there is an open ball of $D$ contained in $S$),
and the proof will use the notion of (topological)
cone, which simply means the union of all segments from a given point $V$ to each
point of a given set $A$: this is called the cone of $A$ with vertex $V$.

\noindent
{\bf Proof:} The geometrical situation leads to the cone, given by
$\de{p,q} \mapsto p\ \pi + \de{1-p}\rho\de{q}$, $p \in \De{0,1}$. The condition on the
tangent vector (together with the fact that $\pi$ is interior to
$S$, while $\rho\de{q}$ has no point in this interior) is  sufficient for this cone
to be $C^m$, except at the vertex $\pi$, at least locally in $q$.

As $S$ is bounded and convex, and $\pi$ is in its interior, every
straight line from $\pi$ crosses $\partial S$ exactly once. As
$\rho\de{q}$ has no point in the interior of $S$, this crossing
always happens for $0 \leq p < 1$. Denote this crossing value by
$p_c\de{q}$. The curve $q \mapsto p_c\de{q}\pi +
\de{1-p_c\de{q}}\rho\de{q}$ is $C^m$, implying $p_c$ is a $C^m$
function of $q$.

The random robustness is given by $R_R\de{\rho\de{q}} =
\frac{p_c}{1-p_c}$. As $p_c < 1$, we also obtain that $R_R$ is a
$C^m$ function of $q$. $\square$

We insist on the interpretation: Proposition \ref{teo} means that any singularity
in $R_R$ for a well choosen path $\rho\de{q}$ reflects singularities in $\partial S$.

From this point on, we study the situation for two qubits, which
is related to the performed experiment described here. In this
case, Ref.~\cite{BraVian} shows that the Random Robustness is
proportional to the Negativity (${\cal N}(\rho)$)\cite{Vidal},
given by the absolute value of the sum of the negative eigenvalues
of the partial transposed state. The negativity is a monotone
under local operations and classical communication \cite{Jens} and
has the operational interpretation of a cost function under a
certain class of operations \cite{Koen, Isi}.

At the same time, entanglement can be measured with the help of
entanglement witnesses \cite{Barbie}. These are Hermitian
operators with positive mean value for all separable states, but
with a negative mean value for some entangled states
\cite{wit1-wit2}. In fact, many geometrical entanglement
quantifiers are directly related to witness operators \cite{Bra}.
In the particular case of two qubits~\footnote{An optimal
entanglement witness $W_{opt}$ satisfying \eqref{Ew} is
proportional to the partial transposition of the projector over
the eigenspace of the negative eigenvalue of $\rho^{T_2}$, where
$\rho^{T_2}$ denotes the partial transposition of $\rho$
\cite{LKCH}.}, we have that for every entangled state $\rho$
\cite{BraVian},
\begin{equation}\label{Ew}
2{\cal N}(\rho) = R_R(\rho) = - 2\min_{W \in {\cal W}} \tr(W \rho),
\end{equation}
where ${\cal W}$ is the set of entanglement witnesses $W$ with $\tr{W} =
2$.

At this point we might ask some natural questions. Is there in fact any singularity in the shape of $S$? In
the affirmative case, does this singularity appear in any physical setup? We
proceed to answer positively both questions by showing physical processes where a singularity in $\partial S$ is revealed by monitoring
the entanglement of a given system.

First, let us consider a general system of four qubits $a$, $b$, $A$, and $B$, subject to
the following Hamiltonian \cite{swap}:
\begin{equation}\label{hamilt1}
H=H^{aA}+H^{bB},
\end{equation}
where
\begin{equation}\label{HaA}
H^{\mu\nu}=\frac{\omega}{2}\sigma_{z}^{\mu}+
\frac{\omega}{2}\sigma_{z}^{\nu}+
\frac{g}{2}(\sigma_{-}^{\mu}\sigma_{+}^{\nu}+\sigma_{+}^{\mu}\sigma_{-}^{\nu}).
\end{equation}
Here $\sigma_{+}=(\sigma_x +i\sigma_y)/2$ and
$\sigma_{-}=(\sigma_x -i\sigma_y)/2$, where
$\sigma_x$, $\sigma_y$ and $\sigma_z$ are the usual Pauli matrices.
This scenario can be realized in many systems, like cavity QED \cite{CQED},
trapped ions \cite{Ions}, and quantum dots \cite{dots}. We set the initial
state to be
$\ket{\psi(t=0)}=\ket{\Phi_{+}}_{ab}\otimes\ket{\Psi_{+}}_{AB}$,
 where qubits $ab$ are in the Bell state
$\ket{\Phi_+}=(\ket{00}+\ket{11})/\sqrt{2}$ and qubits $AB$ are in
the orthogonal Bell state
$\ket{\Psi_{+}}=(\ket{01}+\ket{10})/\sqrt{2}$. Hamiltonian
\eqref{hamilt1} induces a swapping process
which leads (in the interaction picture) to the following temporal
evolution for the subsystem $AB$, obtained by tracing out the
subsystem $ab$:
\begin{equation}\label{AB}
\rho_{AB}(t)= q
\ket{\Psi_{+}}\bra{\Psi_{+}}+ \de{1-q} \ket{\Phi_{+}}\bra{\Phi_{+}},
\end{equation}
where $q=\cos^2(gt)$.
For this state the negativity reads
\begin{equation}\label{RrAB}
{\cal N}(\rho_{AB}(t))=\max\DE{1-2q,2q-1} = \abs{1-2q}.
\end{equation}
This function presents a singularity for
$q=0.5$ ($gt=n\pi/4$, with $n$ odd) signaling then a singularity
at $\partial S$.

Another physical process which also produces the family of states
\eqref{AB} is the following simple quantum communication task:
Alice prepares a Bell state $\ket{\Phi_+}$ and sends one qubit to
Bob through a quantum channel; if this channel has a probability
$q$ of introducing a bit flip, and $1-q$ of no error at all, the
state \eqref{AB} is the  output of the process \footnote{The
simplest way of drawing the complete line represented by
Eq.~\eqref{AB} is to consider three different initial conditions:
from $\ket{\Phi_+}$ one obtains $q \in [0,1/2)$, from
$\ket{\Psi_+}$,  $q \in (1/2,1]$, and $q = 1/2$ is a fixed point
of this dynamical system.}.

\begin{figure}
\includegraphics[scale=.29]{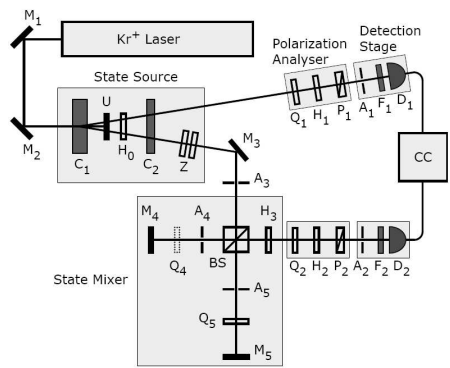}
\caption{{\bf{Experimental setup:}} The state source is composed by
a 2mm-thick BBO ($\beta$-BaB$_2$O$_4$) nonlinear crystal (C$_1$)
pumped by a cw krypton laser operating at 413nm, generating photon
pairs at 826nm by type II spontaneous parametric down-conversion.
Crystal C$_1$ is cut and oriented to generate either one of the
polarization entangled Bell states $|\Psi_{-}\rangle$ or
$|\Psi_{+}\rangle$. Walk-off and phase compensation is provided by
the half-wave plate H$_0$ followed by a 1mm-thick BBO crystal
(C$_{2}$) \cite{SPDC}, together with two 1mm-thick crystalline
quartz plates (Z) inserted in one of the down-converted photon
paths. The unconverted laser beam transmitted by crystal C$_{1}$ is
discarded by means of a dichroic mirror (U).  The detection stages
are composed by photon counting diode modules D$_{1}$ and D$_{2}$,
preceded by 8nm FWHM interference filters F$_{1}$ and F$_{2}$
centered at 825nm, and by circular apertures A$_{1}$ of
1.6mm$\emptyset$ and A$_{2}$ of 3.0mm $\emptyset$. Single and
coincidence counts with 5ns resolving time are registered by a
computer controlled electronic module (CC). Polarization analyzers
are composed by quarter-wave plates Q$_{1}$ and Q$_{2}$, half-wave
plates H$_{1}$ and H$_{2}$, followed by polarizing cubes P$_{1}$ and
P$_{2}$. The State Source produces state $|\Psi_{-}\rangle $. For
each pair, the photon emerging in the upper path goes straight to
the polarization analyzer and to the detection stage $1$. The lower
path photon is directed by mirror M$_{3}$ through the circular
aperture A$_{3}$ into the state mixer (an unbalanced Michelson
interferometer), composed by the beam splitter BS, mirrors M$_{4}$
and M$_{5}$, quarter-wave plates Q$_{4}$ and Q$_{5}$, variable
circular apertures A$_{4}$ and A$_{5}$, and by the half-wave plate
H$_{3}$, whose purpose is to compensate for an unwanted slight
polarization rotation caused by the beam splitter. The quarter-wave
plate Q$_{4}$ is
 switched off which means that if the lower photon follows path labeled $4$, there is no change to its polarization and the half-wave plate
 H$_{3}$ changes the state to $|\Psi_{+}\rangle $. On the other
 hand, if the lower photon follows path labeled $5$, Q$_{5}$ is oriented with the fast axis at 45$^{\circ}$ in order to flip its
 polarization. The path length difference, 130mm, is much larger than the coherence length of the down-converted fields, ensuring an incoherent
 recombination at BS. The pair detected by CC is in state $q |\Psi_{+}\rangle \langle\Psi_{+}| + (1-q) |\Phi_{+}\rangle \langle \Phi_{+}|$
 where $q$ is defined by the relative sizes of apertures A$_{4}$ and A$_{5}$.}
\label{setup}
\end{figure}

\begin{figure}\centering\begin{tabular}{cccc}
    \includegraphics[scale=0.23]{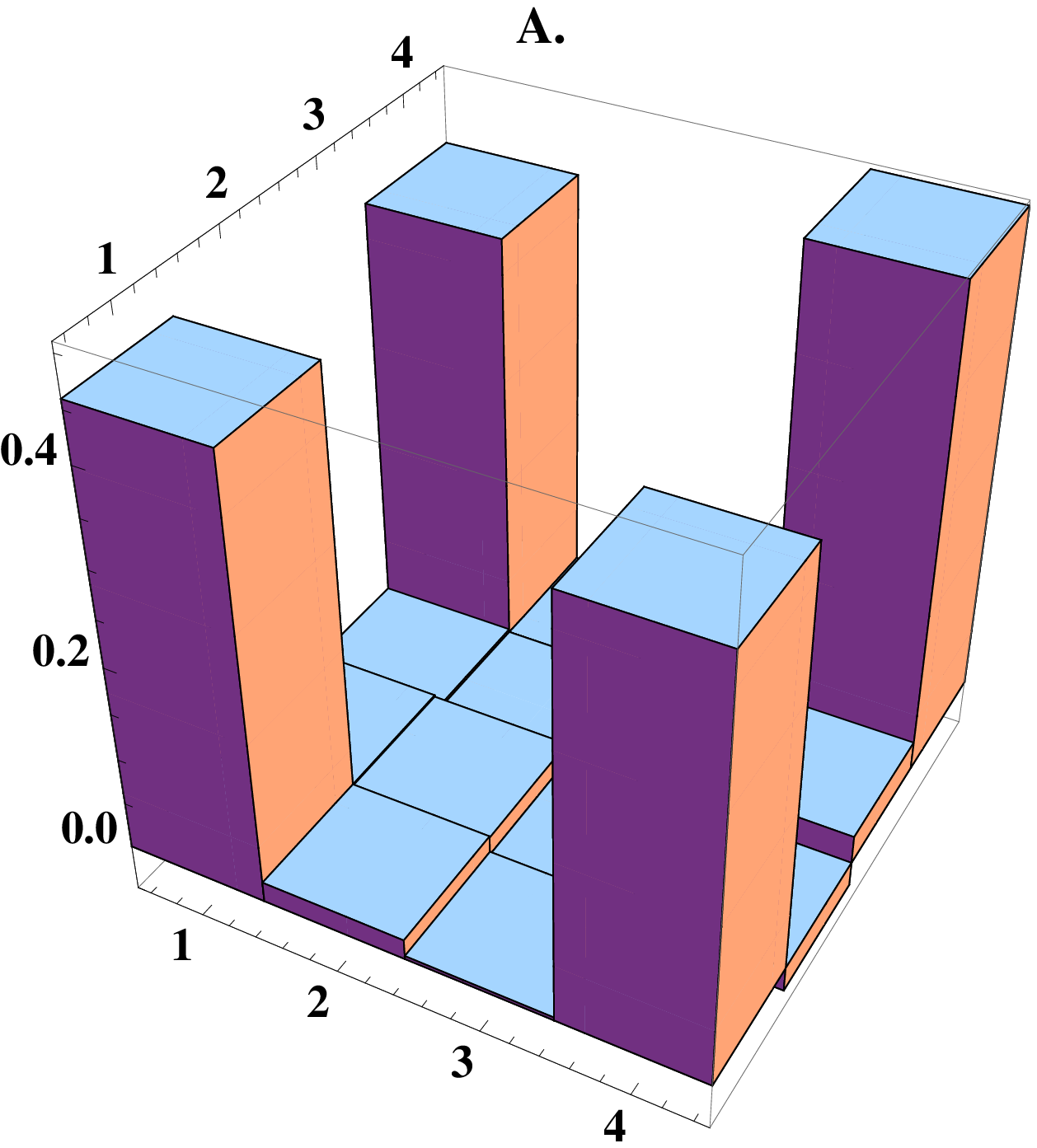} &  \includegraphics[scale=0.23]{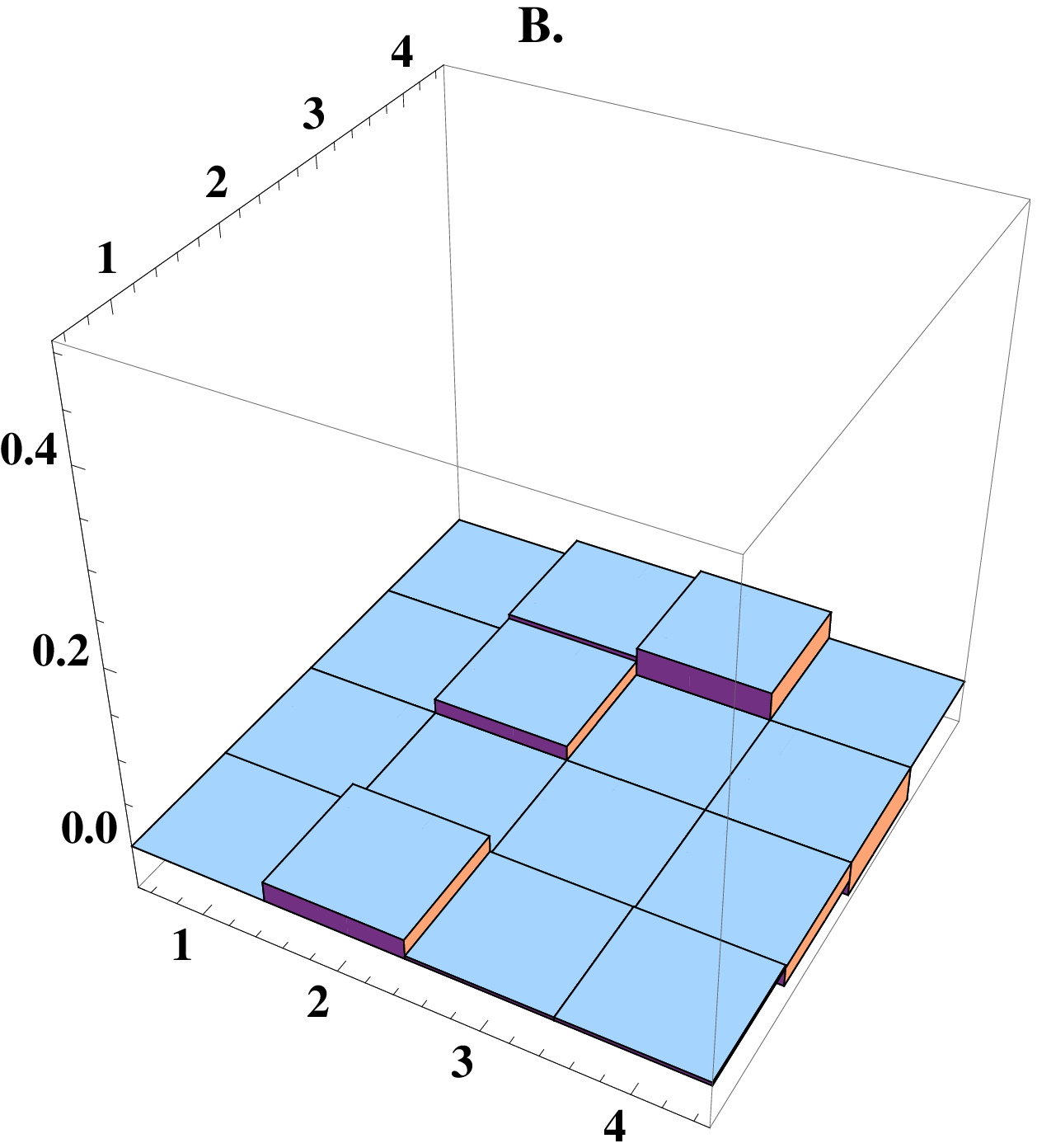} \\
    \includegraphics[scale=0.23]{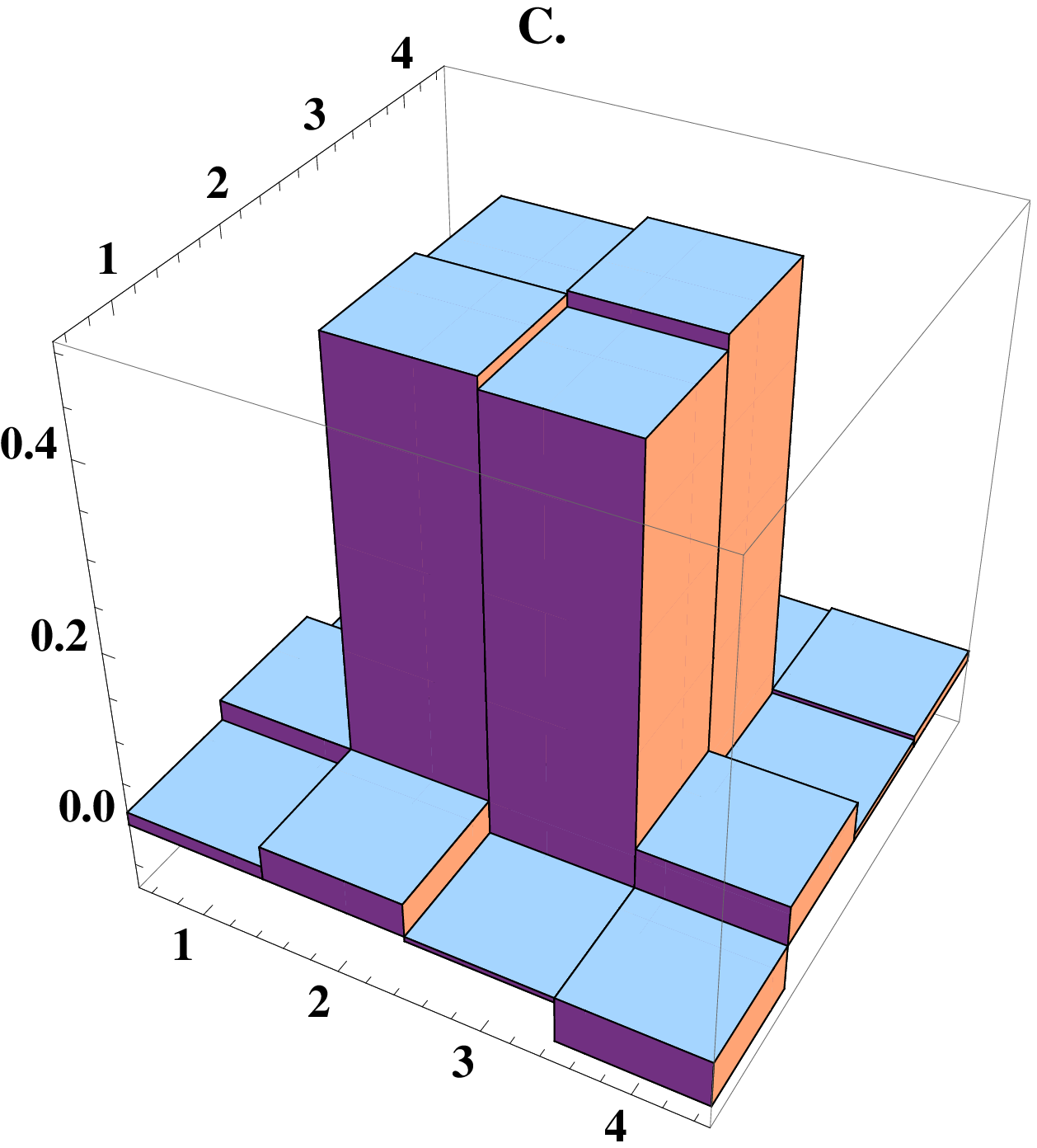} &  \includegraphics[scale=0.23]{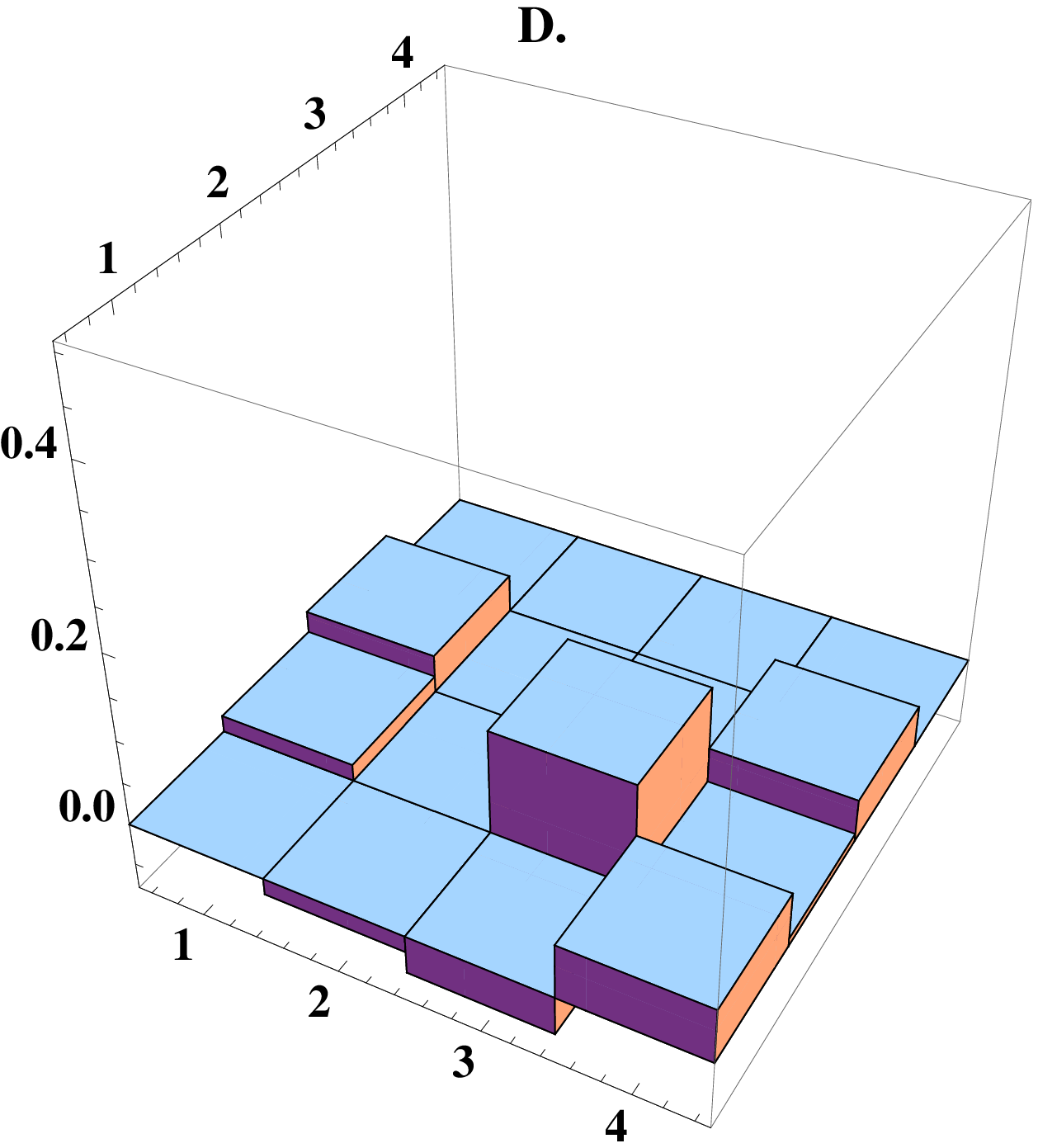} \\
\end{tabular}
   \caption{(Color online) The reconstructed density matrices corresponding (ideally)
   to the states $\ket{\Phi_+}$ (A. real and B. imaginary parts) and $\ket{\Psi_+}$ (C. real and D. imaginary parts).
   The attained fidelity for these states are, respectively, $F_{\Phi_+}\equiv\bra{\Phi_+}\rho\ket{\Phi_+}\approx(92\pm 3)\%$ and $F_{\Psi_+}\equiv\bra{\Psi_+}\rho\ket{\Psi_+}\approx(96\pm 3)\%$.}
\label{Tomo}\end{figure}

\begin{figure}\centering
  \includegraphics[scale=0.28]{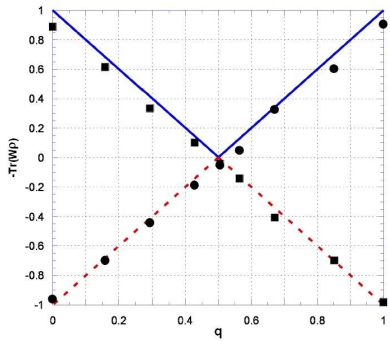} 
\caption{(Color online) Measurement of the mean value of both
operators described in ~\eqref{wit} for the full range $0 \leq q
\leq 1$. Each $W$ is expanded as a linear combination of products
of local operators which are then measured independently. The blue
continuous line corresponds to the theoretical value of
${\mathcal{N}}\de{\rho\de{q}}$ for the state
$\rho(q)=q\ket{\Psi_+}\bra{\Psi_+}+(1-q)\ket{\Phi_+}\bra{\Phi_+}$.
Note that each $W$ only witnesses entanglement for a restricted
range of $q$ values as predicted by the theory. T he local
singularity of $\partial S$ is evidenced by the abrupt change of
optimal $W$. Experimental errors are within the dots' sizes.
}\label{medidas}
\end{figure}

To illustrate the dynamics given by Eq.~\eqref{AB}, we have
performed an optical experiment, shown in Fig.~\ref{setup}. In our
experiment, twin photons maximally entangled in polarization are
generated in a non-linear crystal \cite{SPDC} and sent to an
unbalanced Michelson interferometer, which is used to simulate the
channel described above. The experiment works as follows: we
produce a two-photon $\ket{\Psi_+}$ state, send one of the photons
directly to the detection stage, and the other to the (unbalanced)
interferometer. One of the arms of this interferometer does not
change the polarization of the photon, and if the photon went
through this path the two photons would be detected in
$\ket{\Psi_+}$. However if the photon went through the other path
its polarization would be rotated in such a way that the final
two-photon state would become $\ket{\Phi_+}$. We have made a
tomographic characterization of the photonic states corresponding
to these two extremal points. The reconstructed density matrices
are displayed in Fig. \ref{Tomo}. These two possibilities are then
incoherently recombined, thus allowing the preparation of state
\eqref{AB}. Each preparation yields a different value for $q$ with
the corresponding optimal witness given by
\begin{equation}
\label{wit}
\begin{array}{c}
 W_{opt}=\left\{\begin{array}{rl}
 I-2\ket{\Phi_+}\bra{\Phi_+}\ ,& {\text{ for }} 0\leq q\leq 1/2,\\
 I-2\ket{\Psi_+}\bra{\Psi_+}\ ,& {\text{ for }} 1/2\leq q\leq 1.
 \end{array}\right.
 \end{array}
\end{equation}
For the family of generated states these two observables are the
only candidates of optimal entanglement witnesses, so they are the
only ones to be measured. In a more general situation, if less is
known about the prepared state, much more candidate witnesses
should be measured. The results are displayed in
Fig.~\ref{medidas}. The blue curve in the figure shows the
witnessed negativity measurement and its edge indicates the
existence of singularities at $\partial S$. This experimental
result shows the abrupt change in the optimal witness at the value
$q = \frac{1}{2}$, which heralds the singularity in $\partial S$.
As a proof of principles, each operator $W$ is measured for the
whole range of $q$, which yields the points bellow zero in
Fig.~\ref{medidas}. Note that the singularity occurs exactly for
$R_R = 0$ ($q=1/2$). According to our geometrical interpretation,
this means the path followed by the parameterized state $\rho(q)$
touches the border of $S$. This result must not be a surprise,
since it is well known that in the tetrahedron generated by the
Bell states (which we access in our experiment) the separable
states form a inscribed octahedron \cite{HH}.

The geometrical properties of entanglement discussed here give new
insight into singularities found recently in the entanglement of
condensed matter systems. Striking examples, dealing with
entanglement properties of certain spin-$\frac{1}{2}$ models
subjected to a transverse magnetic field $h$, are described in
Refs.~\cite{Osterloh-Roscilde1-Roscilde2}. In these works, the
two-qubit reduced state shows a singularity in entanglement for a
particular field value $h_f$ far from the critical field of the
respective model. As correlation functions, ground state energy, and
even reduced density matrices are all smooth at $h_f$, there was no
clear origin for these singularities. Our results offer an
explanation by interpreting the non-analyticities exhibited by
entanglement as a consequence of geometric singularities at
$\partial S$~\footnote{Although the results of
Refs.~\cite{Osterloh-Roscilde1-Roscilde2} were obtained in terms of
the concurrence, a completely analogous result holds for the
negativity as well.}.

As previously mentioned, $R_R$ can be used to probe $\partial S$
in any finite dimensional system. For example, a previous work
showed a singular behavior of $R_R$ in three qubits
systems~\cite{BraVian}. Within the scope of our paper, we can
interpret it as originated by a singularity at the border of the
respective separable set. Note, however, that in this case, due to
the higher dimensionality of the system, the singularity at
$\partial S$ occurs in the interior of $D$, with $R_R$ showing a singularity
at a positive value.

To sum up, we have presented a method for probing the shape of the
set of separable states. Singularities in this set were found and
connected to non-analytical behavior of entanglement in different
physical systems. It is an interesting open question to find physical
implications of such singularities.

\begin{acknowledgements} We acknowledge discussions with A. Ac\'in, A.
Sen(De), J. Wehr, E. Rico, G. Palacios, V. Vedral, and J.
Dunningham and funding from CNPq, Fapemig, PRPq-UFMG, and
Brazilian Millenium Institute for Quantum Information.

\end{acknowledgements}

\end{document}